\documentclass{jpsj-suppl}
\usepackage{txfonts} 

\title{Finite-volume Hamiltonian method for $\pi\pi$ scattering in lattice QCD}

\author{Jia-Jun \textsc{ Wu}$^{1}$, T.-S. Harry \textsc{Lee}$^{2}$, Derek B. \textsc{Leinweber}$^{1}$,  A. W. \textsc{Thomas}$^{1}$, and Ross D. \textsc{Young}$^{1}$}

\inst{$^{1}$Special Research Center for the Subatomic Structure of Matter (CSSM),
School of Chemistry and Physics, University of Adelaide Adelaide 5005, Australia \\
$^{2}$Physics Division, Argonne National Laboratory, Argonne, Illinois 60439, USA}

\email{jiajun.wu@adelaide.edu.au}

\recdate{August 21, 2015}

\abst{Within a formulation of $\pi\pi$ scattering, we investigate the use of the finite-volume Hamiltonian
approach to resolving scattering observables from lattice QCD spectra. 
We consider spectra in the centre-of-mass and moving frames for both S- and P-wave cases.
Furthermore, we investigate the multi-channel case. 
Here we study the use of the Hamiltonian framework as a parametrization that can be fit directly to lattice spectra. 
Through this method, the hadron properties, such as mass, width and coupling, can be directly extracted
from the lattice spectra. }

\kword{Lattice, hadron spectrum, Hamiltonian }

\begin{document}
\maketitle

\section{Introduction}

Lattice QCD (LQCD) studies are making significant progress in determining the excitation spectra of hadron. 
%
The L\"uscher method is the established way to connect the spectrum of eigenstates from LQCD and the phase shift and inelastic factor of the S matrix from experiment~\cite{Luscher1,Luscher2}.
However, this method is hard to apply directly in multi-channel cases. 
%
For example, for two coupled channels, there are three asymptotic scattering parameters (i.e., two phase shifts and an inelasticity) corresponding to a single energy eigenstate on a given volume. 
As a result, if all of these three parameters are solved by the L\"uscher equation, it is necessary to search for near-coincident energy eigenstates at three different volumes or through different momentum boosts of the system \cite{guo}.
%

In the present work, we extend a recently developed finite-volume Hamiltonian formalism \cite{hall, wu} in the $\pi\pi$ scattering system.
In the Hamiltonian formalism, the eigenvalues of the Hamiltonian matrix in the finite volume correspond to the eigenstate spectrum of LQCD.
On the other hand, the partial-wave S matrix can be calculated from the Hamiltonian in the infinite volume.
Therefore, it not only provides a bridge between the LQCD spectrum and S-matrix parameters, but it also allows an analysis of them separately.
In this paper, various cases of the $\pi\pi$ system are investigated in the Hamiltonian formalism.
Numerically, the equivalence with the L\"uscher formalism is established for each case.
Furthermore, the pole position of the $\rho$ meson is extracted from recent LQCD data~\cite{Jlabrho}. 

\section{Hamiltonian Model for $\pi\pi$ scattering}

The Hamiltonian is divided as non-interacting and interacting parts: $H = H_0 + H_I$. The $H_0$ is:
\begin{eqnarray}
H_0 =\sum_{i} |\sigma_i\rangle\,\ m^{0}_i\,\langle\sigma_i|
+ \sum_{\alpha} \int d\vec{k}\ |\alpha(\vec{k})\rangle\,\left[\sqrt{m_{\alpha_1}^2 + \vec{k}^{\,\,2}} +
\sqrt{m_{\alpha_2}^2 + \vec{k}^{\,\,2}}\right]\,\langle\alpha(\vec{k})|,
\label{eq:h0}
\end{eqnarray}

where $\sigma_i$ is the $i$-th bare particle with mass $ m^{0}_i$,
$\alpha = \pi\pi, K\bar{K}, \pi\eta, \cdot\cdot$ denotes the channels
included, and $m_{\alpha_i}$ and $\vec{k}$ are the mass and
the momentum of the $i$-th particle in the channel $\alpha$,
respectively.

Following Refs.\cite{mat, kamano}, the $H_I$ includes vertex interactions ($g$) and
two-body potentials ($v$):
\begin{eqnarray}
H_I =g+v = \sum_{\alpha}\int d\vec{k} \sum_{i=1,n} \left\{ |\sigma_i\rangle\,g_{i,\alpha}(k)\,\langle \alpha(\vec{k})|+h.c.\right\}+ \sum_{\alpha,\beta} \int d\vec{k}\,d\vec{k}'\,|\alpha(\vec{k})\rangle  v_{\alpha\,\beta}(k,k')\,\langle \beta(\vec{k}')|.
\end{eqnarray}

From this Hamiltonian it is straightforward to compute the S matrix as shown in Eqs. (6-12) of Ref.\cite{wu}. As a result, the phase shifts and inelasticity obtained from the S matrix are able to be obtained from the Hamiltonian model with the parameters of the interacting parts $H_I$.

%
%

\section{Hamiltonian Matrix in the Finite-Volume}

In this section, the Hamiltonian matrix in the finite-volume is introduced in the rest and moving frames with different partial waves. Furthermore, the eigenvalues of the Hamiltonian matrix solved from $\det[EI-H] = 0$ can be directly compared with the eigenstate spectra of LQCD.

\subsection{S-Wave $\&$ CM system}

In a box, the momenta of the channels are discrete as $\vec{k}_n=(2\pi/L)\,\vec{n}$ where $\vec{n} \in \mathbb{Z}^3$. Correspondingly, the continuum states $|\alpha(\vec{k})>$ are replaced by discrete states $(2\pi/L)^{3/2}|\alpha,\vec{k}_n>$, and the normalization is $\langle \alpha,\vec{k}_n|\beta,\vec{k}_m\rangle  = \delta_{\alpha\beta}\delta_{n_xm_x}\delta_{n_ym_y}\delta_{n_zm_z}$. As a result, the new discrete Hamiltonian is:
\begin{equation}
H_0=\sum_{i} |\sigma_i\rangle m^{0}_i \langle\sigma_i|
+ \sum_{\alpha}\sum_{\vec{n} \in \mathbb{Z}^3} |\alpha,\vec{k}_n\rangle\,\left[\sqrt{m_{\alpha_1}^2 + \vec{k}_n^{\,\,2}} +
\sqrt{m_{\alpha_2}^2 + \vec{k}_n^{\,\,2}}\right]\,\langle\alpha,\vec{k}_n|,
\label{eq:h0cm}
\end{equation}
\begin{equation}
H_I=g+v =\left(2\pi/L\right)^{\frac{3}{2}} \sum_{i}\sum_{\alpha}\sum_{\vec{n} \in \mathbb{Z}^3}\left\{ 
|\sigma_i\rangle\,g_{i,\alpha}\,\langle \alpha,\vec{k}_n|+h.c.\right\}
+\left(2\pi/L\right)^{3}  \sum_{\alpha,\beta}\sum_{\vec{n}, \vec{m}\in \mathbb{Z}^3} |\alpha,\vec{k}_n\rangle\,  v_{\alpha\,\beta}\,\langle \beta, \vec{k}_m|.
\label{eq:gvcm}
\end{equation}
The eigenvalues, $E$, of the Hamiltonian matrix are solved from $\det[EI-H_0-g-v] = 0$.   For the application, only the main points are listed here since very detailed discussions are in Ref.~\cite{wu}.
\begin{itemize}
\item  We have demonstrated the equivalence of the Hamiltonian finite volume spectra with the L\"uscher formalism for both a single channel and the corresponding generalization to a coupled-channel system.
\item Beyond the inelastic threshold, the finite-volume Hamiltonian offers a much more tractable approach for extracting the S-matrix parameters from the finite-volume spectra.
\item  Through the fitting approach, the Hamiltonian matrix provides a robust description of hadronic interactions from lattice  spectra. The results are independent of the model within the fitted energy region.
\end{itemize}
 
\subsection{S-Wave $\&$ Boosted system}

In this section, the total momentum, $\vec{P}$, of the system is non-zero. Thus, the basic state  $|\alpha,\vec{k}_n>$ becomes $|\alpha,\vec{P},\vec{k}_n>$. 
Because the interaction part of Hamiltonian is only defined in the center-of-mass system, the whole Hamiltonian is calculated as the center-of-mass energy. 
Furthermore,  it requires an additional factor
$C_\alpha(\vec{k},\vec{P})=\sqrt{\frac{\left(E_{\alpha_1}(\vec{k})+E_{\alpha_2}(\vec{P}-\vec{k})\right)E_{\alpha_1}(\vec{k}^*)E_{\alpha_2}(\vec{k}^*)}{\left(E_{\alpha_1}(\vec{k}^*)+E_{\alpha_2}(\vec{k}^*)\right)E_{\alpha_1}(\vec{k})E_{\alpha_2}(\vec{P}-\vec{k})}}$
 which is the square root of the Jacobi factor between the momentum in the center-of-mass system, $\vec{k}^*$, and the moving system, $\vec{k}$.
 As a result, the new Hamiltonian is:
\begin{eqnarray}
&&H_0 = \sum_{i} |\sigma_i\rangle\,m^{0}_i\,\langle\sigma_i|
+ \sum_{\alpha}\sum_{\vec{n} \in \mathbb{Z}^3} |\alpha,\vec{P},\vec{k}_n\rangle\,\left[\sqrt{\left(E_{\alpha1}(\vec{k}_n) + E_{\alpha2}(\vec{P}-\vec{k}_n)\right)^2-\vec{P}^2}\,\right]\,\langle\alpha,\vec{P},\vec{k}_n|,
\label{eq:h0bs}\\
&&H_I=g+v =\ 
\left(2\pi/L\right)^{\frac{3}{2}}
\sum_{i,\alpha}\sum_{\vec{n} \in \mathbb{Z}^3}C_\alpha(\vec{k}_n,\vec{P})
\left\{
|\sigma_i\rangle\,g_{i,\alpha}\,\langle \alpha,\vec{P},\vec{k}_n|+h.c.
\right\}\nonumber\\
&&\ \ \ \ \ \ \ \ \ \ \ \ +
\left(2\pi/L\right)^{3}
\sum_{\alpha,\beta}\sum_{\vec{n}, \vec{m}\in \mathbb{Z}^3}C_\alpha(\vec{k}_n,\vec{P})\,C_\beta(\vec{k}_m,\vec{P})\,|\alpha,\vec{P},\vec{k}_n\rangle\, v_{\alpha\,\beta}\,\langle \beta,\vec{P}, \vec{k}_m|.
\label{eq:vbs}
\end{eqnarray}

After the parameters of the Hamiltonian are constrained by experimental data, the eigenvalues of the Hamiltonian matrix for different total momentum are calculated. These are shown as black solid lines in Fig. \ref{fg:spec}(a). The spectra (red lines) directly solved by the L\"uscher equation from fitted phase shift are shown as red dash lines. They are very consistent at the large lattice size, L$>$ 3 fm. 


\subsection{P-Wave $\&$ C.M. system and Boost system}

In the above two subsections, the kinematic factor of the Hamiltonian is discussed in the rest and moving frames. Rotational symmetry is trivial because only the S-wave is considered. Higher partial waves are neglected. 
For the P-wave case, the rotational symmetry plays an important role, since the orbit angular momentum L of the $\pi\pi$ states are non-zero.
However, L is not a good quantum number in the moving box. 
The rotational symmetry of the moving box obeys the little group, rather than the SO(3) group. 
As a result, the spectra of the moving box follow the irreducible representations of the little group. 
In Ref.~\cite{meissner}, the coefficients, $C_{\Gamma,\Gamma_ n, L, m}$, between basis state $|\Gamma,\Gamma_n>$ of the little group and the usual $|L,m>$ are given.
In the Hamiltonian model, $H_0$ is the same as before, while the interaction part is: ( for the P-wave of $\pi\pi$ scattering, neglecting the $v$ potential)
\begin{eqnarray}
&&g = \left(2\pi/L\right)^{\frac{3}{2}}
\sum_{\alpha}\sum_{\vec{n} \in \mathbb{Z}^3}
C_\alpha(\vec{k}_n,\vec{P})
\sum_{\Gamma,\Gamma_n}
\left\{  
|\rho, \Gamma,\Gamma_ n\rangle\,g_{\rho\pi\pi}(|\vec{k}^*|,\Gamma,\Gamma_n)\,\langle \alpha,\vec{P},\vec{k}_n|+h.c.\right\},
\label{eq:gbs}\\
&& g_{\rho\pi\pi}(|\vec{k}^*|,\Gamma,\Gamma_n)=\sum_{m} C_{\Gamma,\Gamma_n,1,m}\,g_{\rho\pi\pi}(|\vec{k}^*|,1,m)=u(|\vec{k}^*|)\,\sum_{m} C_{\Gamma,\Gamma_n,1,m}\,\epsilon_\mu(\vec{P},m)\,k^\mu ,
\end{eqnarray}
where $u(|\vec{k}^*|)$ is the form factor defined in the rest frame and $\epsilon_\mu(\vec{P},m)$ is the polarization vector of the $\rho$ meson with momentum $\vec{P}$ and z-component of spin $m$. $k$ is the four-momentum of $\pi$ in the moving system. Clearly, the interactions rely on the representation $\Gamma$, which results in the different spectra of Hamiltonian matrix for the various little groups.

\begin{figure}[tbh]
\includegraphics[width=0.49\columnwidth]{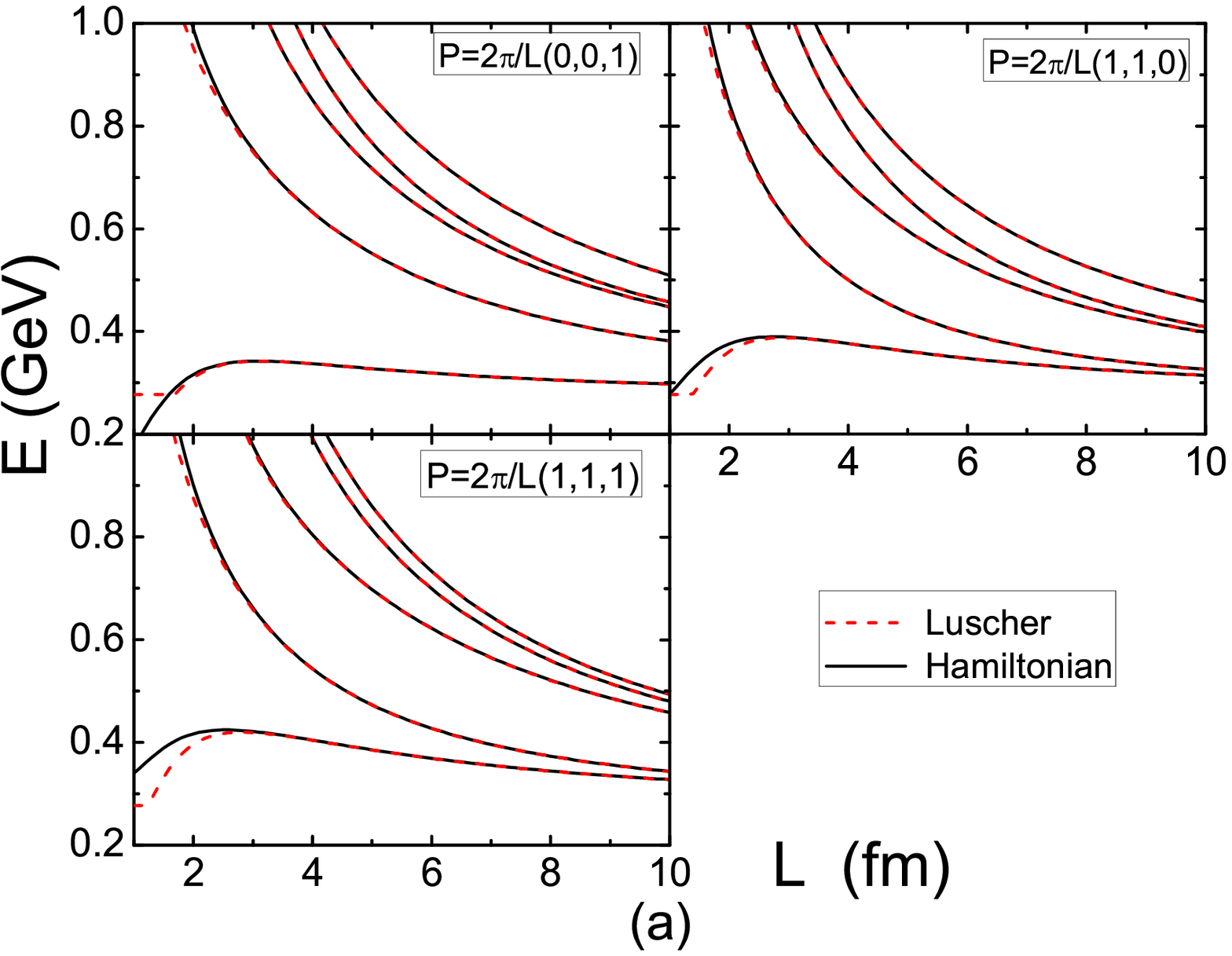}
\includegraphics[width=0.49\columnwidth]{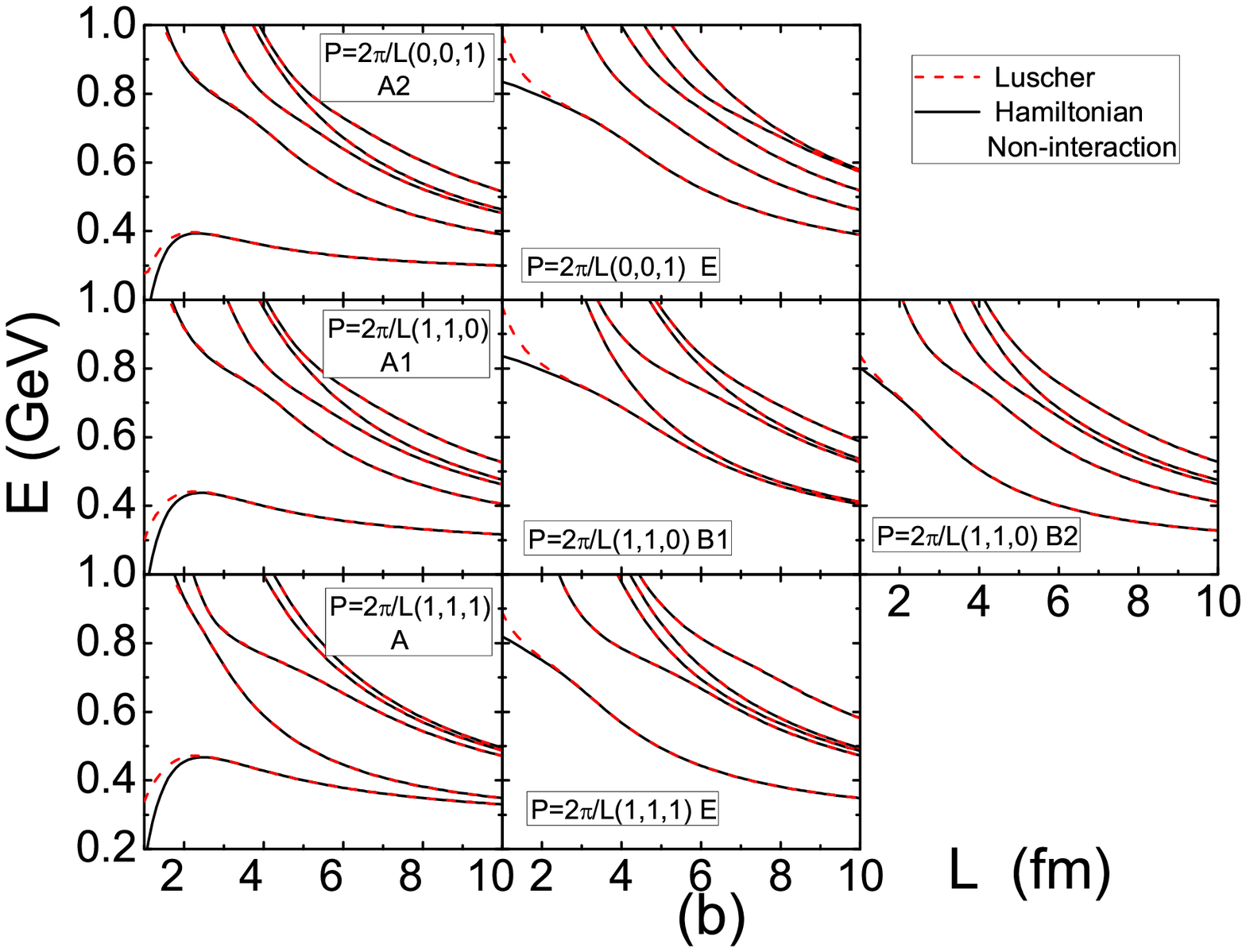}
\includegraphics[width=0.49\columnwidth]{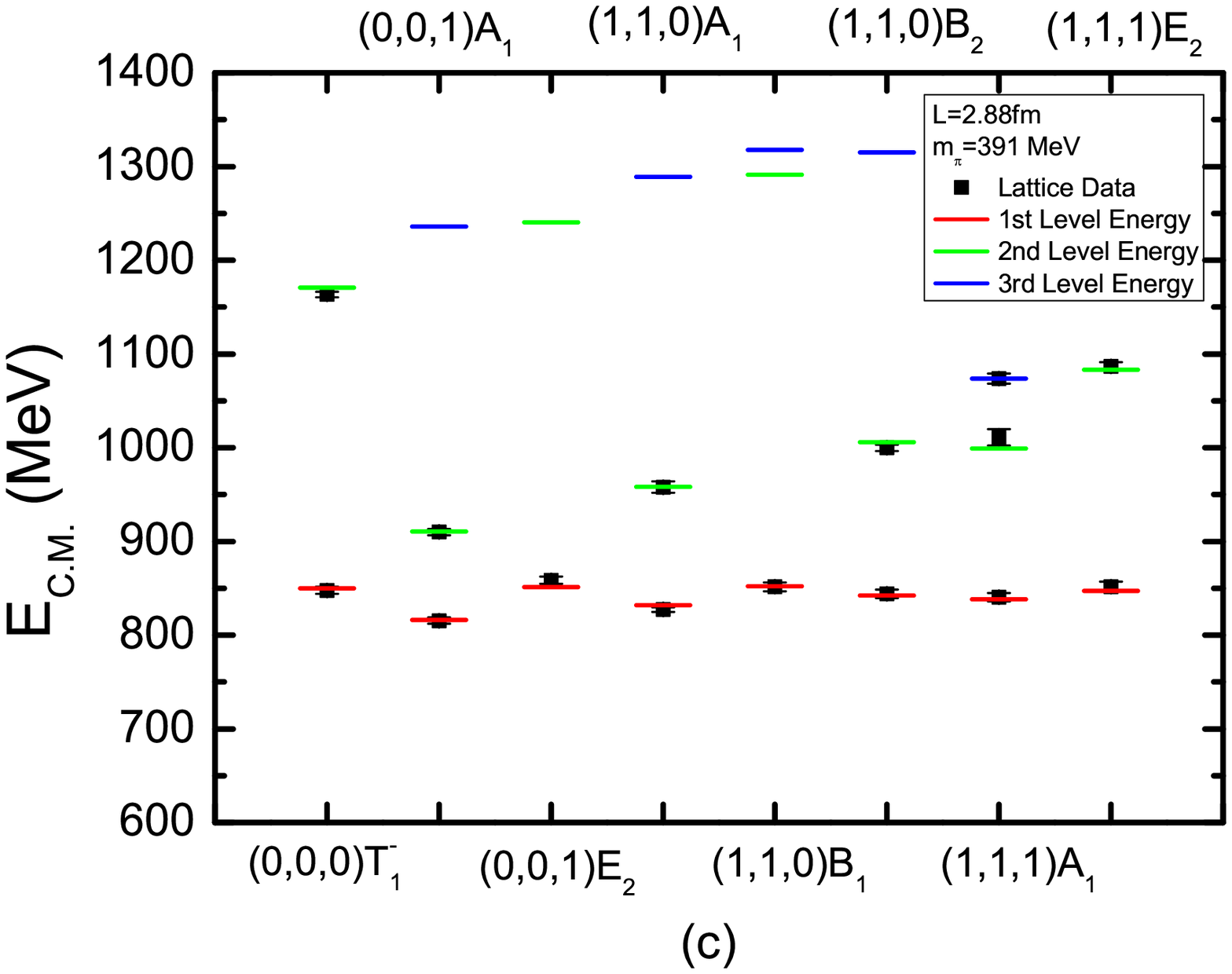}
\caption{ (a) and (b) show the spectra for S-wave and P-wave scattering in moving frames with different total momenta, respectively. (c) Matrix Hamiltonian results are fit to recent Lattice data~\cite{Jlabrho}.}
\label{fg:spec}
\end{figure}

In Fig.~\ref{fg:spec}(b), the spectra (black solid lines) of various total momentum and representations are shown. 
They are very consistent with the results (red dashed lines) solved directly through the L\"uscher equation from phase shifts produced by the same Hamiltonian. 
Furthermore, as shown in Fig.~\ref{fg:spec}(c), the Hamiltonian model fits the recent LQCD data well.~\cite{Jlabrho} Then $\rho$ meson pole position is: $m-i\frac{\Gamma}{2} = 840.5-i5.0$ MeV. The Breit-Wigner parameters are $m=854.1$ MeV and $\Gamma = 12.1$ MeV, determined by using a K matrix model adapted to the  L\"uscher formalism~\cite{Jlabrho}.
 It is notable that there is still 14 and 2 MeV difference between mass and width, respectively. The difference is typical of that between the pole position in the complex energy plane and Breit-Wigner definition on the real axis.

\section{Summary}

In this paper, the Hamiltonian model for $\pi\pi$ scattering is investigated in both finite and infinite volume. In previous work \cite{wu}, the equivalence to the well established L\"uscher equations has been demonstrated. 
Furthermore, the extension of the Hamiltonian model to th moving system and P-wave is briefly introduced. 
All of these demonstrate that the Hamiltonian method can now be used to calculate any cases of two body spectra in a box. On the other side, the Hamiltonian model offers a robust framework  for the parameterization of hadronic interactions to fit lattice spectra. This provides an alternative method to extract resonance information directly from the LQCD data.  
As an example, by analyzing the new LQCD data from Jlab \cite{Jlabrho}, the corresponding $\rho$-meson resonances properties have been extracted. 
These investigations show that the Hamiltonian approach provides  a new competitive method to the existing L\"uscher formalism. Future work will apply this method to various lattice hadronic spectra to obtain resonance information.

$\\$
\textbf{Acknowledgment}

This work is supported by the University of Adelaide and the
Australian Research Council through the ARC Centre of Excellence for
Particle Physics at the Terascale and grants FL0992247 (AWT),
DP140103067, FT120100821 (RDY).

\end{document}